\documentclass[twocolumn,showpacs,preprintnumbers,amsmath,amssymb]{revtex4}
\usepackage{epsfig}
\usepackage{graphicx}

\usepackage{epsfig}

\begin{document}


\title{Hashing protocol for distilling multipartite CSS states}

\author{Erik Hostens}
\email{erik.hostens@esat.kuleuven.be}
\affiliation{ESAT-SCD, K.U.Leuven, Kasteelpark Arenberg 10, B-3001 Leuven, Belgium}
\author{Jeroen Dehaene}
\affiliation{ESAT-SCD, K.U.Leuven, Kasteelpark Arenberg 10, B-3001 Leuven, Belgium}
\author{Bart De Moor}
\affiliation{ESAT-SCD, K.U.Leuven, Kasteelpark Arenberg 10, B-3001 Leuven, Belgium}
\date{\today}

\newcommand{\T}{{\cal T}_\epsilon^{(k)}}
\newcommand{\Z}{\mathbb{Z}}
\newcommand{\Jt}{{\cal J}^\perp}

\begin{abstract}
We present a hashing protocol for distilling multipartite CSS states by means of local Clifford operations, Pauli measurements and classical communication. It is shown that this hashing protocol outperforms previous versions by exploiting information theory to a full extent an not only applying CNOTs as local Clifford operations. Using the information-theoretical notion of a strongly typical set, we calculate the asymptotic yield of the protocol as the solution of a linear programming problem.
\end{abstract}


\maketitle

\section{Introduction}

Stabilizer states and codes are an important concept in quantum information theory. Stabilizer codes \cite{GPhD,
G:98} play a central role in the theory of quantum error correcting codes, which protect quantum information
against decoherence and without which effective quantum computation has no chance of existing. Recently, a
promising alternative setup for quantum computation has been found that is based on the preparation of a stabilizer state (more specifically a cluster state) and one-qubit measurements \cite{R:03}. Also in the
area of quantum cryptography and quantum communication, both bipartite as multipartite, the number of applications of stabilizer states is abundant. We cite Refs.~\cite{dur2,B2,B3,ekert,karlsson,hillery,cleve,crep}, but this is far from an exhaustive list.

Closely related to quantum error correction, entanglement distillation is a means of extracting
entanglement from quantum states that have been disrupted by the environment. Many applications require pure multipartite
entangled states that are shared by remote parties. In practice, these pure states are prepared by one party and
communicated to the others by an imperfect quantum channel. As a result, the states are no longer pure. A
distillation protocol then consists of local operations combined with classical communication in order to end up
with states that approach purity and are suited for the application in mind. An interesting distillation
protocol for Bell states is the well-known hashing protocol, introduced in Ref.~\cite{Bennett}, that has its roots in
classical information theory.

In this paper, we describe a generalization of this hashing protocol from bipartite to multipartite.
It distills an important particular kind of stabilizer states, called CSS states, short for Calderbank-Shor-Steane states. Bell states, cat states and cluster
states (more generally two-colorable graph states) are examples of or locally equivalent to CSS states. In
brief, the protocol goes as follows: $k$ copies of an $n$-qubit mixed state are shared by $n$ remote parties.
They perform local unitary operations and measurements that, if $k$ is large, result in a state that approaches $\gamma k$ copies of a pure $n$-qubit CSS state, where $\gamma<1$ is the yield of the protocol. The basic idea of
describing the protocol in a classical information theoretical setting is the same as in Ref.~\cite{Bennett}.

Very similar multipartite hashing protocols have been discussed in Refs.~\cite{dur1,asch}, Ref.~\cite{man} and Ref.~\cite{lo} for two-colorable graph states, cat states and CSS states respectively. Our protocol improves these protocols in two ways. First, we note that in Refs.~\cite{dur1,asch,man,lo}, by not exploiting information theory to a full extent, their protocols result in overkill. In short, demanding that the number of measurements exceed the marginal entropies \cite{dur1,asch,man} of each separate party results in too many measurements. In Ref.~\cite{lo}, this is partially meeted by relaxing to conditional entropies. We will show that our protocol is optimal in the given setting and is therefore a complete generalization of the hashing protocol for Bell states to CSS states. The yield is calculated as the solution of a linear programming problem, and requires a somewhat more involved information-theoretical treatment. A second major difference is that the local unitary operations applied in Refs.~\cite{dur1,asch,man,lo} only consist of CNOTs, whereas in some cases a higher yield can be achieved by using more general local Clifford operations. To this end, we need to know which local Clifford operations result in a permutation of all possible $2^{nk}$ $k$-fold tensor products of an $n$-qubit CSS state. This is done efficiently using the binary matrix description of stabilizer states and Clifford operations of Ref.~\cite{D:03}.

This paper is organized as follows. In section~\ref{sectionbinary}, we introduce the binary framework in which
stabilizer states and Clifford operations are efficiently described. In section~\ref{sectioninfo}, we define the
strongly typical set, an information-theoretical concept that is needed to calculate the yield. In
section~\ref{sectionperm}, we derive necessary and sufficient conditions that local Clifford operations have to
satisfy to result in a permutation of the $2^{nk}$ $k$-fold tensor products of an $n$-qubit
CSS state. This result is a generalization of Ref.~\cite{DVD:03}, and is also interesting for more recurrence-like protocols as also introduced in Ref.~\cite{dur1,asch}. But we will not go deeper into this issue in this paper. In section~\ref{sectionprotocol}, we explain how our hashing protocol works and calculate
the yield in section~\ref{sectionyield}. Finally, the protocol is illustrated and compared to others by an example in section~\ref{sectionex}. Readers that are merely interested in the results can skip almost entirely sections~\ref{sectioninfo}, \ref{sectionperm}, \ref{sectionyield} and the appendices.

\section{Preliminaries}

\subsection{Stabilizer states, CSS states and Clifford operations in the binary picture}\label{sectionbinary}

In this section, we present the binary matrix description of stabilizer states and Clifford operations. We show how Clifford operations act on stabilizer states in the binary picture. We also formulate a simple criterion for separability of a stabilizer state. CSS states are then defined as a special kind of stabilizer states, and we show the particular properties of their binary matrix description. We will restrict ourselves to definitions and properties that are necessary to the distillation protocols presented in the next sections. In the following, all addition and multiplication is performed modulo 2. For a more elaborate discussion on the binary matrix description of stabilizer states and Clifford operations, we refer to Ref.~\cite{D:03}.

We use the following notation for Pauli matrices.
\[\begin{array}{ccccl}\sigma_{00} &=& I_2 &=& \left[\begin{array}{rr}1&0\\0&1\end{array}\right],\\
\sigma_{01} &=& \sigma_x &=& \left[\begin{array}{rr}0&1\\1&0\end{array}\right],\\
\sigma_{10} &=& \sigma_z &=& \left[\begin{array}{rr}1&0\\0&-1\end{array}\right],\\
\sigma_{11} &=& \sigma_y &=& \left[\begin{array}{rr}0&-i\\i&0\end{array}\right].\end{array}\]
Let $v,w\in\Z_2^n$ and $a=\left[\begin{array}{cc}v\\w\end{array}\right]$, then we denote
\begin{equation*}
\sigma_a=\sigma_{v_1w_1}\otimes\ldots\otimes\sigma_{v_nw_n}.
\end{equation*}
The Pauli group on $n$ qubits is defined to contain all tensor products $\sigma_a$ of Pauli matrices with an additional complex phase factor in $\{1,i,-1,-i\}$. In this paper we will only consider Hermitian Pauli operators, so we may exclude imaginary phase factors. Note that all Hermitian Pauli operators square to the identity. It can also be easily verified that Pauli operators satisfy the following commutation relation:
\begin{equation}\label{comm}
\sigma_a\sigma_b=(-1)^{a^TPb}\sigma_b\sigma_a,~\mbox{where}~P=\left[\begin{array}{cc}0&I_n\\I_n&0\end{array}\right].
\end{equation}

A stabilizer state $|\psi\rangle$ on $n$ qubits is the simultaneous eigenvector, with eigenvalues 1, of $n$ commuting Hermitian Pauli operators $(-1)^{b_i}\sigma_{s_i}$, where $s_i\in\Z_2^{2n}$ are linearly independent and $b_i\in\Z_2$, for $i=1,\ldots, n$. The $n$ Hermitian Pauli operators generate an Abelian subgroup of the Pauli group on $n$ qubits, called the stabilizer $\cal S$. We will assemble the vectors $s_i$ as the columns of a matrix $S\in\Z_2^{2n\times n}$ and the bits $b_i$ in a vector $b\in\Z_2^n$. Note that it follows from (\ref{comm}) that commutativity of the stabilizer is reflected by $S^TPS=0$. The representation of $\cal S$ by $S$ and $b$ is not unique, as every other generating set of $\cal S$ yields an equivalent description. In the binary picture, a change from one generating set to another is represented by an invertible linear transformation $R\in\Z_2^{n\times n}$ acting on the right on $S$ and acting appropriately on $b$. We have
\begin{equation}\label{R}\begin{array}{ccl}
S' &=& SR \\
b' &=& R^Tb+d\end{array}\end{equation}
where $d\in\Z_2^n$ is a function of $S$ and $R$ but not of $b$ \cite{D:03}. We will show below that in the context of distillation protocols, $d$ can always be made zero.

Each $S$ defines a total of $2^n$ orthogonal stabilizer states, one for each $b\in\Z_2^n$. As a consequence, all stabilizer states defined by $S$ constitute a basis for ${\cal H}^{\otimes n}$, where $\cal H$ is the Hilbert space of one qubit. In the following, we will refer to this basis as the $S$-basis.

A Clifford operation $Q$, by definition, maps the Pauli group to itself under conjugation:
\[Q\sigma_a Q^{\dag}=(-1)^{\delta}\sigma_b.\]
It is clear that the Pauli group is a subgroup of the Clifford group, as
\[\sigma_v\sigma_a\sigma_v^\dag=(-1)^{v^TPa}\sigma_a.\]
In the binary picture, a Clifford operation is represented by a matrix $C\in\Z_2^{2n\times 2n}$ and a vector $h\in\Z_2^{2n}$, where $C$ is symplectic or $C^TPC=P$ \cite{D:03}. The image of a Hermitian Pauli operator $\sigma_{a}$ under the action of a Clifford operation is then given by $(-1)^{\epsilon}\sigma_{Ca}$, where $\epsilon$ is function of $C,h$ and $a$. Note that the phase factor of the image can always be altered by taking $Q'=Q\sigma_g$ instead of $Q$, where $\sigma_g$ anticommutes with $\sigma_a$, or $a^TPg=1$, as
\[Q'\sigma_a {Q'}^{\dag}=Q\sigma_g\sigma_a\sigma_g^{\dag}Q^{\dag}=-Q\sigma_a Q^{\dag}.\]
If a stabilizer state $|\psi\rangle$, represented by $S$ and $b$, is operated on by a Clifford operation $Q$, represented by $C$ and $h$, $Q|\psi\rangle$ is a new stabilizer state whose stabilizer is given by $Q{\cal S}Q^{\dag}$. As a result, this stabilizer is represented by
\begin{equation}\label{CS}\begin{array}{ccl}
S' &=& CS \\
b' &=& b+f\end{array}\end{equation}
where $f$ is independent of $b$ and can always be made zero, by performing an extra Pauli operator $\sigma_g$ before  the Clifford operation, where $S^TPg=f$. Because $S$ is full rank, this equation always has a solution. The resulting Clifford operation is then $Q'=Q\sigma_g$ instead of $Q$. With this, $C$ remains the same, but $b'=b$ in (\ref{CS}). In the same way, $d$ in (\ref{R}) can be made zero. Thus, from now on, we may neglect the influence of $h$ on the protocol and represent a Clifford operation only by $C$.

Let $|\psi_1\rangle$ and $|\psi_2\rangle$ be two stabilizer states represented by $S_1=\left[\begin{array}{c}S_{1(z)}\\S_{1(x)}\end{array}\right], b_1$ and $S_2=\left[\begin{array}{c}S_{2(z)}\\S_{2(x)}\end{array}\right], b_2$ respectively. Then $|\psi_1\rangle\otimes|\psi_2\rangle$ is a stabilizer state represented by
\begin{equation}\label{stabprod}
\left[\begin{array}{cc}S_{1(z)} & 0\\0 & S_{2(z)}\\S_{1(x)} & 0\\0 & S_{2(x)}\end{array}\right],~\left[\begin{array}{cc}b_1\\b_2\end{array}\right].\end{equation}
Conversely, a stabilizer state $|\psi\rangle$ represented by $S,b$ is separable iff there exists a permutation matrix $T\in\Z_2^{n\times n}$ and an invertible matrix $R\in\Z_2^{n\times n}$ such that $(I_2\otimes T)SR$ has a block structure as in (\ref{stabprod}). Note that left multiplication with $(I_2\otimes T)$ on $S$ is equivalent to permuting the qubits and right multiplication with $R$ on $S$ yields another representation of $|\psi\rangle$.

Let $Q_1$ and $Q_2$ be two Clifford operations represented by $\left[\begin{array}{cc}A_1&B_1\\C_1&D_1\end{array}\right]$ and $\left[\begin{array}{cc}A_2&B_2\\C_2&D_2\end{array}\right]$ respectively, where all blocks are in $\Z_2^{n\times n}$. Then $Q_1\otimes Q_2$ is a Clifford operation represented by
\begin{equation}\label{clifprod}
\left[\begin{array}{cccc}A_1 & 0 & B_1 & 0\\0 & A_2 & 0 & B_2\\C_1 & 0 & D_1 & 0\\0 & C_2 & 0 & D_2\end{array}\right].\end{equation}

A CSS state, or Calderbank-Shor-Steane state, is a stabilizer state $|\psi\rangle$ whose stabilizer can be represented by
\begin{equation}\label{CSS}
S=\left[\begin{array}{cc}S_z & 0\\0 & S_x\end{array}\right],~b
\end{equation}
where $S_z\in\Z_2^{n\times n_z}$, $S_x\in\Z_2^{n\times n_x}$ and $n_z+n_x=n$. The stabilizer condition $S^TPS=0$ is equivalent to $S_z^TS_x=0$. As $S$ is full rank, $S_z$ and $S_x$ are also full rank. Therefore, once $S_z$ (or $S_x$) is known, we know $S$, up to right multiplication with some $R$. The following statements involving $S_z$ also hold when using $S_x$. The state $|\psi\rangle$ is separable iff there exists a permutation matrix $T\in\Z_2^{n\times n}$ and an invertible matrix $R\in\Z_2^{n_z\times n_z}$ such that
\[TS_zR=\left[\begin{array}{cc}S_z' & 0\\0 & S_z''\end{array}\right],\]
where $S_z'\in\Z_2^{n'\times n_z'}$, $S_z''\in\Z_2^{n''\times n_z''}$, $n'+n''=n$, $n_z'+n_z''=n_z$ and $0<n'<n$.
Indeed, since $S_z'$ and $S_z''$ are full rank, it is possible to find $S_x'\in\Z_2^{n'\times(n'-n_z')}$ and $S_x''\in\Z_2^{n''\times(n''-n_z'')}$ such that ${S_z'}^TS_x'=0$ and ${S_z''}^TS_x''=0$. The stabilizer that results from the qubit permutation $T$ is represented by
\[\left[\begin{array}{cccc}S_z' & 0 & 0 & 0\\0 & 0 & S_z'' & 0\\0 & S_x' & 0 & 0\\0 & 0 & 0 & S_x''\end{array}\right]\]
which has the block structure defined in (\ref{stabprod}).

If the phase factors $(-1)^{b_i}$, for $i=1,\ldots, n$, of a CSS state represented by (\ref{CSS}) are unknown, a $\sigma_z$ measurement on every qubit reveals $b_i$, for $i=1,\ldots, n_z$. Indeed, the measurements project the state on the joint eigenspace of observables $\sigma_z^{(j)}=I_2^{\otimes j-1}\otimes\sigma_z\otimes I_2^{\otimes n-j}$, for $j=1,\ldots, n$, with eigenvalues $(-1)^{a_{j}}$ that are determined by the measurements. We then have
\[b=\left[\begin{array}{c}S_z^Ta \\ \ast \end{array}\right].\]
The last $n_x$ phase factors $\ast$ are lost due to the fact that all $\sigma_{s_i}$, for $i=n_z+1,\ldots, n$, anticommute with at least one $\sigma_z^{(j)}$. On the other hand, by $\sigma_x$ measurements on every qubit, with outcomes $(-1)^{a_j}$, we learn that
\[b=\left[\begin{array}{c}\ast \\ S_x^Ta \end{array}\right].\]
More generally, we can divide $\{1,\ldots,n\}$ into two disjunct subsets $M_z$ and $M_x$. A $\sigma_z$ measurement on every qubit $i\in M_z$ and a $\sigma_x$ measurement on every qubit $i\in M_x$ reveals all $r^Tb$, $r\in\Z_2^n$, for which $Sr$ has zeros on positions $i$ for $i\in M_x$ and on positions $n+i$ for $i\in M_z$.

\subsection{Strongly typical set}\label{sectioninfo}

In this section, we introduce the information-theoretical notion of a \emph{strongly typical set}. We will need this in section~\ref{sectionyield}. This section is self-contained, but for an introduction to information theory, we refer to Ref.~\cite{CT}.

Let $X=(X_1,\ldots,X_k)$ be a sequence of independent and identically distributed discrete random variables, each having event set $\Omega$ with probability function $p: \Omega\mapsto [0,1]: a\mapsto p(a)$. The strongly typical set $\T$ is defined to be the set of sequences $x=(x_1,\ldots,x_k)\in\Omega^k$ for which the sample frequencies $f_a(x)=|\{x_i~|~x_i=a\}|/k$ are close to the true values $p(a)$, or: $x\in\T~\Leftrightarrow$
\begin{equation}\label{defT}
|f_a(x)-p(a)|<\epsilon,~\forall a\in\Omega.
\end{equation}
It can be verified that $f_a(X)$ has mean $p(a)$ and variance $p(a)[1-p(a)]/k$. By Chebyshev's inequality \cite{cheb}, we have
\[P(|f_a(X)-p(a)|\geq\epsilon)\leq \frac{p(a)[1-p(a)]}{k\epsilon^2}.\]
It follows that $p(\T)\geq 1-\delta$, where $\delta=O(k^{-1}\epsilon^{-2})$.

In section~\ref{sectionyield}, we will encounter the following problem. Let $\Omega$ be partitioned into subsets $\Omega_j$ ($j=1,\ldots, q$). We define the function
\[y(x)=(\Omega_{j_1},\ldots,\Omega_{j_k}),~\mbox{where}~x_i\in\Omega_{j_i},~\mbox{for}~i=1,\ldots, k.\]
Given some $u\in\T$, calculate the number $|{\cal N}_u|$ of sequences $v\in\T$ that satisfy $y(v)=y(u)$, or
\[{\cal N}_u=\{v\in\T~|~y(v)=y(u)\}.\]
For all $v\in {\cal N}_u$ and for $j=1,\ldots, q$, it holds
\begin{equation}\label{f}
\sum_{a\in\Omega_j}f_a(v)=f_{\Omega_j}(v)=f_{\Omega_j}(u)=\sum_{a\in\Omega_j}f_a(u).
\end{equation}
Fix $f_a$ satisfying (\ref{defT}) and (\ref{f}) and call ${\cal N}_f$ the set of elements $v\in {\cal N}_u$ with these sample frequencies $f_a$. Then elementary combinatorics tells us
\[|{\cal N}_f|=\prod_{j=1}^q\frac{[f_{\Omega_j}(v)k]!}{\prod_{a\in\Omega_j}[f_a(v)k]!}.\]
Using Stirling's approximation \cite{stir} for large $k$:
\[\ln k! = k\ln k -k+O(\ln k),\]
and (\ref{f}) we find that $\log_2 |{\cal N}_f|= O(\log_2 k)+$
\[k\sum_{j=1}^q\left[f_{\Omega_j}(v)\log_2f_{\Omega_j}(v) - \sum_{a\in\Omega_j}f_a(v)\log_2f_a(v)\right].\]
As $v\in\T$, we have that $f_a(v)=p(a)+O(\epsilon)$, for all $a\in\Omega$. Therefore,
\[\log_2 |{\cal N}_f|= k[H(X)-H(Y)+O(\epsilon)]+O(\log_2 k)\]
where $H(X)=-\sum_a p(a)\log_2 p(a)$ is the entropy of $X$ and $H(Y)=-\sum_j p(\Omega_j)\log_2 p(\Omega_j)$ the entropy of $y(X)$. It is clear that $|{\cal N}_f|\leq|{\cal N}_u|$. Since there is a total $\leq (2\epsilon k)^q$ of $f$ that satisfy (\ref{defT}), an upper bound for $|{\cal N}_u|$ is
\[(2\epsilon k)^q\max_f|{\cal N}_f|,\]
where the maximum is taken over all $f$ that satisfy (\ref{defT})-(\ref{f}). It follows that
\begin{equation*}
|{\cal N}_u|=2^{k[H(X)-H(Y)+O(\epsilon)]+O(\log_2k)}.
\end{equation*}

\section{Local permutations of products of CSS states}\label{sectionperm}

In this section, we consider $n$-qubit CSS states that are all represented by the same $S$. We have $k$ states that are shared by $n$ remote parties, each holding corresponding qubits of all $k$ states. We study local Clifford operations (local with respect to the partition into $n$ parties) that result in a permutation of all $2^{nk}$ possible tensor products of such CSS states. As the distillation protocol described in the next section only consists of local operations, we may assume that $S$ defines fully entangled states. Indeed, if $S$ would define separable states, the protocol would be two simultaneous protocols that do not influence each other.

If $|\psi_i\rangle$ ($i=1,\ldots, k$) are represented by
\begin{equation*}
S=\left[\begin{array}{cc}S_z & 0\\0 & S_x\end{array}\right],~b_i
\end{equation*}
according to (\ref{stabprod}), $|\psi_1\rangle\otimes\ldots\otimes|\psi_k\rangle$ is represented by
\begin{equation*}
\left[\begin{array}{cc}I_k\otimes S_z & 0\\0 & I_k\otimes S_x\end{array}\right],~\tilde{b}'=\left[\begin{array}{c}b_1\\\vdots\\b_k\end{array}\right].
\end{equation*}
However, since it is more convenient to arrange all qubits per party, we rewrite the stabilizer matrix by permuting rows and columns as
\begin{equation}\label{copies}
\left[\begin{array}{cc}S_z\otimes I_k & 0\\0 & S_x\otimes I_k\end{array}\right]=S\otimes I_k,~\tilde{b}
\end{equation}
where the entries of $\tilde{b}'$ are permuted appropriately into $\tilde{b}\in\Z_2^{nk}$. All parties perform local Clifford operations. According to (\ref{clifprod}), the overall Clifford operation is then most generally represented by
\begin{equation}\label{LC}
\left[\begin{array}{cc}\tilde{A} & \tilde{B}\\\tilde{C} & \tilde{D}\end{array}\right]=
\left[\begin{array}{ccc|ccc}
A_1 & & & B_1 & & \\
 & \ddots & & & \ddots & \\
 & & A_n & & & B_n\\ \hline
C_1 & & & D_1 & & \\
 & \ddots & & & \ddots & \\
 & & C_n & & & D_n
\end{array}\right],\end{equation}
where the representations of the local Clifford operations $\left[\begin{array}{cc}A_i & B_i\\C_i & D_i\end{array}\right]\in\Z_2^{2k\times 2k}$ are symplectic matrices, or
\begin{equation}\label{clifcond}\begin{array}{rcl}
A_i^TC_i + C_i^TA_i &=& 0\\
B_i^TD_i + D_i^TB_i &=& 0\\
A_i^TD_i + C_i^TB_i &=& I_k
\end{array}\quad \mbox{for}~i=1,\ldots, n.\end{equation}

The local Clifford operations acting on the given state result in a permutation of all $2^{nk}$ possible tensor products (defined by $\tilde{b}$) iff the resulting stabilizer matrix can be transformed into the original form of (\ref{copies}) by multiplication with an invertible $R\in\Z_2^{nk\times nk}$ on the right, or
\begin{equation}\label{CSR}
\left[\begin{array}{cc}\tilde{A} & \tilde{B}\\\tilde{C} & \tilde{D}\end{array}\right] (S \otimes I_k) R =  S \otimes I_k.
\end{equation}
Using (\ref{R}) and (\ref{CS}), the corresponding permutation of the tensor products is then defined by the transformation
\begin{equation}\label{permutation}
\tilde{b}\mapsto R^T\tilde{b}.
\end{equation}
We now investigate for which local Clifford operations an $R$ can be found such that (\ref{CSR}) holds. Without loss of generality, we may assume that
\begin{equation}\label{theta}
S_z=\left[\begin{array}{c}I_{n_z}\\\theta\end{array}\right],~S_x=\left[\begin{array}{c}\theta^T\\I_{n_x}\end{array}\right]
\end{equation}
where $\theta\in\Z_2^{n_x\times n_z}$. This can be obtained by multiplication with an invertible $R$ on the right. Let
\[\tilde{A}_z=\left[\begin{array}{ccc}
A_1 & & \\
 & \ddots & \\
 & & A_{n_z}
\end{array}\right],~
\tilde{A}_x=\left[\begin{array}{ccc}
A_{n_z+1} & & \\
 & \ddots & \\
 & & A_{n}
\end{array}\right].\]
Using analogous definitions for $\tilde{B}_z,\tilde{B}_x,\tilde{C}_z,\tilde{C}_x,\tilde{D}_z$ and $\tilde{D}_x$, the left hand side of (\ref{CSR}) becomes
\begin{equation*}\begin{array}{c}
\left[\begin{array}{cccc}
\tilde{A}_z & 0 & \tilde{B}_z & 0\\
0 & \tilde{A}_x & 0 & \tilde{B}_x \\
\tilde{C}_z & 0 & \tilde{D}_z & 0\\
0 & \tilde{C}_x & 0 & \tilde{D}_x
\end{array}\right]
\left[\begin{array}{cc}
I_{n_z}\otimes I_k & 0\\
\theta\otimes I_k & 0\\
0 & \theta^T\otimes I_k \\
0 & I_{n_x}\otimes I_k
\end{array}\right] R = \\
\left[\begin{array}{cc}
\tilde{A}_z & \tilde{B}_z(\theta^T\otimes I_k) \\
\tilde{A}_x(\theta\otimes I_k) & \tilde{B}_x \\
\tilde{C}_z & \tilde{D}_z(\theta^T\otimes I_k) \\
\tilde{C}_x(\theta\otimes I_k) & \tilde{D}_x
\end{array}\right] R.
\end{array}\end{equation*}
We can now write (\ref{CSR}) as two separate equations:
\begin{eqnarray}
\nonumber\left[\begin{array}{cc}
\tilde{A}_z & \tilde{B}_z(\theta^T\otimes I_k) \\
\tilde{C}_x(\theta\otimes I_k) & \tilde{D}_x
\end{array}\right] R &=& I_{nk} \\
\nonumber\left[\begin{array}{cc}
\tilde{C}_z & \tilde{D}_z(\theta^T\otimes I_k) \\
\tilde{A}_x(\theta\otimes I_k) & \tilde{B}_x
\end{array}\right] R &=&
\left[\begin{array}{cc}
0 & \theta^T\otimes I_k \\
\theta\otimes I_k & 0
\end{array}\right].\\
\label{Req} & &
\end{eqnarray}
\vspace{-6mm}

\noindent Eliminating $R$, we get
\[\begin{array}{c}
\left[\begin{array}{cc}
0 & \theta^T\otimes I_k \\
\theta\otimes I_k & 0
\end{array}\right]
\left[\begin{array}{cc}
\tilde{A}_z & \tilde{B}_z(\theta^T\otimes I_k) \\
\tilde{C}_x(\theta\otimes I_k) & \tilde{D}_x
\end{array}\right] = \\
\left[\begin{array}{cc}
\tilde{C}_z & \tilde{D}_z(\theta^T\otimes I_k) \\
\tilde{A}_x(\theta\otimes I_k) & \tilde{B}_x
\end{array}\right],\end{array}\]
which is a necessary and sufficient condition on the local Clifford operations (\ref{LC}) such that an $R$ exists that satisfies (\ref{CSR}). Blockwise comparison of both sides yields the following equations
\begin{eqnarray}
\label{A}(\theta\otimes I_k)\tilde{A}_z &=& \tilde{A}_x(\theta\otimes I_k) \\
\label{D}(\theta^T\otimes I_k)\tilde{D}_x &=& \tilde{D}_z(\theta^T\otimes I_k) \\
\label{B}(\theta\otimes I_k)\tilde{B}_z(\theta^T\otimes I_k) &=& \tilde{B}_x \\
\label{C}(\theta^T\otimes I_k)\tilde{C}_x(\theta\otimes I_k) &=& \tilde{C}_z
\end{eqnarray}
From (\ref{A})-(\ref{D}) and the fact that $\theta$ represents fully entangled CSS states, it follows that (see Appendix~\ref{appclif})
\begin{equation}\label{IA}
\begin{array}{cccccccc}
A_1 &=& \ldots &=& A_n & \equiv & A\\
D_1 &=& \ldots &=& D_n & \equiv & D.
\end{array}\end{equation}

Furthermore, if $\theta$ is orthogonal, or $\theta^T\theta=I_{n/2}$ where $n$ is even, it follows from (\ref{B})-(\ref{C}) that the same holds for $B_i$ and $C_i$. Thus, we have
\[\left[\begin{array}{cc}
\tilde{A} & \tilde{B} \\
\tilde{C} & \tilde{D}
\end{array}\right] =
\left[\begin{array}{cc}
I_n\otimes A & I_n\otimes B \\
I_n\otimes C & I_n\otimes D
\end{array}\right].\]
If $\theta$ is orthogonal, then $S_z^TS_z=0$ and it is better to represent the stabilizer by choosing $S_x=S_z$ instead of (\ref{theta}). With this, the left hand side of (\ref{CSR}) becomes
\[\left[\begin{array}{cc}
I_n\otimes A & I_n\otimes B \\
I_n\otimes C & I_n\otimes D
\end{array}\right]
\left[\begin{array}{cc}
S_z\otimes I_k & 0 \\
0 & S_z\otimes I_k
\end{array}\right]R\]
which, with (\ref{clifcond}), is equal to $S\otimes I_k$ iff
\begin{equation}\label{RO}
R=\left[\begin{array}{cc}
I_{n/2}\otimes D^T & I_{n/2}\otimes B^T \\
I_{n/2}\otimes C^T & I_{n/2}\otimes A^T
\end{array}\right].\end{equation}

However, mostly $\theta$ is not orthogonal. In that case, (\ref{B})-(\ref{C}) can only hold (see Appendix~\ref{appclif}) if $B_i=0$ for all $i\in Z_B$ and $C_i=0$ for all $i\in Z_C$, for some $Z_B,Z_C\subseteq\{1,\ldots,n\}$ and $Z_B\cup Z_C=\{1,\ldots,n\}$. So we always have either $B_i$ or $C_i$ equal to zero, for every $i=1,\ldots, n$. From (\ref{clifcond}) it then follows that $D=(A^T)^{-1}=A^{-T}$ and that $A^TC_i$ and $A^{-1}B_i$ are symmetric, for all $i=1,\ldots, n$. Note that local Clifford operations (\ref{LC}) that satisfy these properties together with (\ref{IA}) form a subgroup of the Clifford group. Only for these local Clifford operations, (\ref{A})-(\ref{C}) hold. With (\ref{Req}), it can now be verified that
\begin{equation}\label{RN}
R=\left[\begin{array}{cc}
I_{n_z}\otimes A^{-1} & \tilde{B}_z^T(\theta^T\otimes I_k) \\
\tilde{C}_x^T(\theta\otimes I_k) & I_{n_x}\otimes A^T
\end{array}\right].
\end{equation}
Finally, we mention that (\ref{B})-(\ref{C}) are equivalent to the following linear constraints (see Appendix~\ref{appclif}):
\begin{eqnarray}
\label{B'}(\left[\begin{array}{cc}\theta & I_{n_x} \\ L_{\theta^T}^T & 0\end{array}\right]\otimes I_k)\left[\begin{array}{c}B_1\\\vdots\\B_n\end{array}\right]&=&0\\
\label{C'}(\left[\begin{array}{cc}I_{n_z} & \theta^T \\ 0 & L_\theta^T\end{array}\right]\otimes I_k)\left[\begin{array}{c}C_1\\\vdots\\C_n\end{array}\right]&=&0.
\end{eqnarray}
The $n_z$-bit columns of $L_{\theta^T}$ are $(\theta^T)_j\odot(\theta^T)_l,~\forall j,l: 1\leq j<l\leq n_x$, which stands for the elementwise product of columns $j$ and $l$ of $\theta^T$. An analogous definition holds for $L_\theta$. This will be of interest in section~\ref{sectionyield}.

Finally, we summarize this section. For a particular CSS state, we want a general formula for $R$ such that (\ref{CSR}) holds. First, we rewrite $S$ in the form of (\ref{theta}). Then we distinguish two cases. If $\theta$ is orthogonal, then $R$ is given by (\ref{RO}). If $\theta$ is not orthogonal, then $R$ is given by (\ref{RN}) where the constraints (\ref{B'})-(\ref{C'}) must be satisfied. Note that the symplecticity condition (\ref{clifcond}) remains to be satisfied at all times.

\section{Protocol}\label{sectionprotocol}

In this section, we show how the hashing protocol for CSS states is carried out. As noted in section~\ref{sectionbinary}, all $2^n$ stabilizer states represented by the same $S\in\Z_2^{2n\times n}$ constitute a basis for ${\cal H}^{\otimes n}$, which we call the $S$-basis. The protocol starts with $k$ identical copies of a mixed state $\rho$ that is diagonal in this basis. This mixed state could for instance be the result of distributing $k$ copies of a pure CSS state, represented by $S$ and $b=0$, via imperfect quantum channels. If $\rho$ is not diagonal in the $S$-basis, it can always be made that way by performing a local POVM. We refer to Ref.~\cite{asch} for a proof. We have
\begin{equation*}
\rho=\sum\limits_{b\in\Z_2^n}p(b)|\psi_b\rangle\langle\psi_b|,
\end{equation*}
where $|\psi_b\rangle$ is the CSS state represented by $S$ and $b$. The mixed state $\rho$ can be regarded as a statistical ensemble of pure states $|\psi_b\rangle$ with probabilities $p(b)$. Consequently, $k$ copies of $\rho$ are an ensemble of pure states represented by (\ref{copies}) with probabilities
\begin{equation}
p(\tilde{b})=p(\tilde{b}')=\prod_{i=1}^{k} p(b_i).
\end{equation}
Recall that the entries of $\tilde{b}$ correspond to the $nk$ phase factors ordered per party instead of per copy like $\tilde{b}'$.

\begin{figure}
\includegraphics[width=0.4\textwidth]{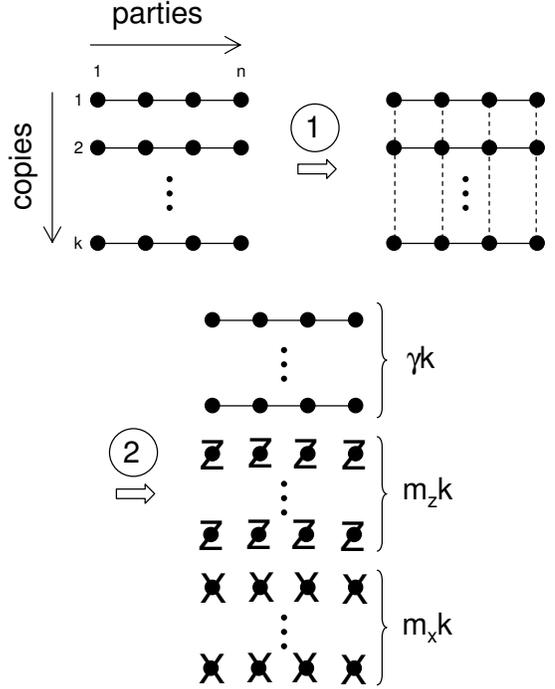}
\caption{\label{protocol} in the first step, local Clifford operations (local with respect to the parties) result in statistically dependent copies. In the second step, some of the copies are measured, providing information on the global state. Afterwards, the measured copies are separable.}
\end{figure}

The protocol now consists of the following steps (this is schematically depicted in figure~\ref{protocol}):
\begin{enumerate}
\item Each party applies local Clifford operations (\ref{LC}) that result in the transformation (\ref{permutation}) of $\tilde{b}$. Consequently, all $2^{nk}$ tensor products represented by the $2^{nk}$ different $\tilde{b}$ in the ensemble are permuted.
\item A fraction $mk$ of all $k$ copies are measured locally. These copies are divided in two sets with $m_zk$ and $m_xk$ copies respectively ($m_z+m_x=m$). Each of the $n$ parties performs a $\sigma_z$ measurement on every qubit they have of the first set of copies, and a $\sigma_x$ measurement on every qubit of the second set.
\end{enumerate}
The local Clifford operations result in a permutation $\tilde{b}\mapsto R^T\tilde{b}$ of all tensor products such that the ensembles of the different copies become statistically dependent. We will specify $R$ later. The measurements provide information on the overall state. The goal of the protocol is to collect enough information for the $(1-m)k$ remaining copies to approach a pure state. The yield $\gamma=1-m$ of the protocol is the fraction of pure states that are distilled out of $k$ copies, if $k$ goes to infinity.

It is important to mention that, next to exclusive $\sigma_z$ or $\sigma_x$ measurements, the qubits of a copy to be measured could be partitioned into two disjunct sets $M_z$ and $M_x$ and measured appropriately. This too will provide information on the state, as explained in section~\ref{sectionbinary}. Then all copies to be measured should be divided into a number of sets: one set for each possible partition ($2^n$ in total). Evidently, not all partitions will be interesting and some of them can be ruled out from the beginning. Otherwise, it will follow from the calculations that no copy should be measured according to those partitions. For simplicity, we will restrict ourselves to the partitions $M_x=\emptyset$ (only $\sigma_z$ measurements) or $M_z=\emptyset$ (only $\sigma_x$ measurements). All derivations still hold in the general case. 

Thus far, we have not specified $R$. The measurement outcomes should contain as much information as possible. Therefore, the outcome probabilities should be uniform. This is achieved as follows. Recall that if $\theta$ is orthogonal, all possible $R$ are of the form (\ref{RO}) with constraints (\ref{clifcond}). If $\theta$ is not orthogonal, all possible $R$ are of the form (\ref{RN}) with constraints (\ref{clifcond}) and (\ref{B'})-(\ref{C'}). We now randomly pick an element of the set of all possible $R$. We will prove in the next section that this yields uniform outcome probabilities.

A way of looking at the ensemble is to regard it as an unknown pure state. The probability that the state is represented by $\tilde{b}$ is then equal to $p(\tilde{b})$. Suppose the unknown pure state is represented by $\tilde{u}$. With probability $\geq 1-\delta$, where $\delta=O(k^{-1}\epsilon^{-2})$, $\tilde{u}$ is contained in the set $\T$, defined as in section~\ref{sectioninfo}. Here, $\Omega$ is the set of all $b\in\Z_2^n$. We now assume that $\tilde{u}\in\T$. After each measurement, we eliminate every $\tilde{b}\in\T$ that is inconsistent with the measurement outcome. The protocol has succeeded if all $\tilde{b}\neq\tilde{u}$ are eliminated from $\T$ and only $\tilde{u}$ is left. Indeed, by the assumption made, at least $\tilde{u}$ must survive this process of elimination. With probability $\leq\delta$, this assumption is false: in that case, the protocol will end up with a state presumed to be represented by some $\tilde{b}\in\T$ but is not, which means that the protocol has failed.

In the next section, we will calculate the yield of the protocol as the solution of the following linear programming problem: $\gamma=1-m$, where $m$ is the solution to
\begin{equation*}\begin{array}{lcl}
\mbox{minimize} & & m=m_z+m_x \\
\mbox{subject to} & & d_z m_z+d_x m_x\geq H-H_{[d_z,d_x]}, \\
\mbox{for all} & & [d_z,d_x]\neq[0,0],\\
& & 0\leq d_z\leq n_z, \\
& & 0\leq d_x\leq n_x.
\end{array}\end{equation*}
$H$ is the entropy of the initial mixed state, or 
\begin{equation*}
H=-\sum\limits_{b\in\Z_2^n}p(b)\log_2 p(b).
\end{equation*}
The calculation of $H_{[d_z,d_x]}$ is more involved. Define the subspace $\Jt=\{w\in\Z_2^n|J^Tw=0\}$ of $\Z_2^n$, where $J$ is a matrix with $n$ rows and defined below. The cosets $\Omega_j$ ($j=1,\ldots,q$) of this subspace constitute a partition of $\Z_2^n$. This partition has entropy 
\begin{equation*}
H_{\Jt} = -\sum\limits_{j=1}^qp(\Omega_j)\log_2 p(\Omega_j).
\end{equation*}
Now $H_{[d_z,d_x]}$ is defined as follows:
\begin{equation*}
\min\limits_{{\cal G}_z,{\cal G}_x}H_{\Jt},
\end{equation*}
where the minimum is taken over all subspaces ${\cal G}_z$ of $\Z_2^{n_z}$ with dimension $n_z-d_z$ and subspaces ${\cal G}_x$ of $\Z_2^{n_x}$ with dimension $n_x-d_x$. The matrix $J$ that defines $\Jt$ is function of ${\cal G}_z$ and ${\cal G}_x$ as follows:
\begin{itemize}
\item \underline{if $\theta$ is orthogonal}:\\
We use the representation where $S_x=S_z$. We have
\begin{equation*}
J = \left[\begin{array}{cccc}
G_z & 0 & 0 & G_x \\
0 & G_z & G_x & 0
\end{array}\right].
\end{equation*}
\item \underline{if $\theta$ is not orthogonal}:\\
Let $M_\theta$ be a matrix whose column space is the orthogonal complement of that of $L_\theta$ and $M_{\theta^T}$ likewise for $L_{\theta^T}$ (for a definition of $L_\theta, L_{\theta^T}$ see the end of section~\ref{sectionperm}). Let $G_z\in\Z_2^{n_z\times(n_z-d_z)}, G_x\in\Z_2^{n_x\times(n_x-d_x)}$ be matrices whose column spaces are ${\cal G}_z, {\cal G}_x$ respectively. Then we have
\begin{equation*}
J = \left[\begin{array}{cccc}
G_z & 0 & 0 & V \\
0 & U & G_x & 0
\end{array}\right].
\end{equation*}
The $n_x$ rows of $U$ are the Kronecker products of the corresponding rows of $\theta G_z$ and $M_\theta$. The $n_z$ rows of $V$ are the Kronecker products of the corresponding rows of $\theta^T G_x$ and $M_{\theta^T}$.
\end{itemize}

\section{Calculating the yield}\label{sectionyield}

This section is organized as follows. In the first subsection we show that the outcome probabilities of each measurement are uniform. This is used to calculate the probability that some $\tilde{b}\neq\tilde{u}$ is not eliminated after all measurements. In the second subsection we then calculate the minimal number of measurements needed to eliminate all $\tilde{b}\neq\tilde{u}$. This is stated as a linear programming problem. We will assume that $\theta$ is not orthogonal. All derivations for the other case are very similar.

\begin{figure}
\includegraphics[width=0.45\textwidth]{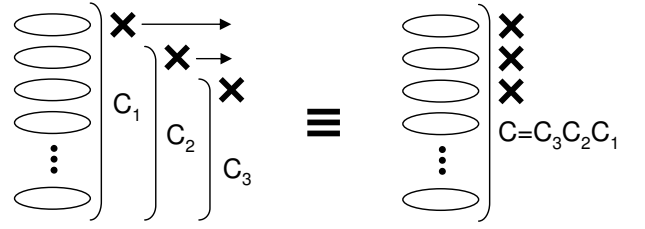}
\caption{\label{eqviews} two equivalent views of the protocol. Subsequent random Clifford operations ($C_1, C_2, C_3$) performed only on non-measured copies, each followed by the measurement of a single copy are equivalent to performing just one random Clifford operation ($C$) and the same measurements.}
\end{figure}

Before we go into the detailed calculation of the yield, we give two different but equivalent views of the protocol. As stated in the previous section, the protocol consists of a Clifford operation followed by measurements. This Clifford operation is randomly picked out of all Clifford operations that are local and result in a permutation as explained in section~\ref{sectionperm}. Now suppose we would perform such a random Clifford operation after every measurement, but only on the copies left (i.e. not measured). As every measurement commutes with every Clifford operation that follows, all measurements can be postponed until the end. It is clear that if all Clifford operations performed are random and yield a permutation, the same holds for the overall Clifford operation. In the following subsection, we will use this second view. Both views are illustrated in figure~\ref{eqviews}.

\subsection{Elimination probability}

We will first calculate the probability that some $\tilde{b}\neq\tilde{u}$ is not eliminated after a $\sigma_z$ measurement on the $i$-th copy. As explained in section~\ref{sectionbinary}, this reveals
\[z_j=(R^T\tilde{u})_{(j-1)k+i},~\mbox{for}~j=1,\ldots, n_z,\]
while
\[x_j=(R^T\tilde{u})_{(n_z+j-1)k+i},~\mbox{for}~j=1,\ldots, n_x,\]
are lost. For a $\sigma_x$ measurement, it is the other way around. Iff $(R)_{(j-1)k+i}^T(\tilde{b}+\tilde{u})=0$ for $j=1,\ldots, n_z$, then $\tilde{b}$ is not eliminated. Assume that the $i$-th copy is the first measured. For the measurement outcome, we are only interested in the $i$-th columns of $A^{-1}$ and $C_l^T$ ($l=n_z+1,\ldots, n$). We define $a=(A^{-1})_i$ and 
\[c=\left[\begin{array}{c}(C_{n_z+1}^T)_i\\\vdots\\(C_n^T)_i\end{array}\right].\]
From the randomness of $R$, it follows that $a$ and $c$ are uniformly distributed over all possibilities. We denote the sets of all possibilities for $a$ and $c$ by ${\cal R}_a$ and ${\cal R}_c$ respectively. It is clear that ${\cal R}_a=\Z_2^{k}\setminus\{0\}$. However, we assume that ${\cal R}_a=\Z_2^{k}$, as there is a negligible probability ($2^{-k}$) that $a$ is chosen equal to $0$ (even during the course of the process, this probability will be $\leq 2^{-\gamma k}$ and $\gamma>0$). From (\ref{C'}), we have
\[{\cal R}_c=\{c\in\Z_2^{n_xk}~|~(L_\theta^T\otimes I_k)c=0\}.\]

We define the matrix $V_z\in\Z_2^{nk\times n_z}$ with columns $(V_z)_j=(R)_{(j-1)k+i}$, for $j=1,\ldots, n_z$, and ${\cal V}_z$ as the set containing all possible values of $V_z$, which is uniformly distributed too. Note that ${\cal V}_z$ is a vector space, because ${\cal R}_a$ and ${\cal R}_c$ are vector spaces and $V_z$ is a linear function of $a$ and $c$. Let $\Delta\tilde{b}=\tilde{b}+\tilde{u}$ and $\Delta z=V_z^T\Delta\tilde{b}$. For some fixed $\Delta\tilde{b}$, all values $\Delta z\in{\cal Z}=\{V_z^T\Delta\tilde{b}~|~V_z\in{\cal V}_z\}$ are equiprobable. Indeed, all cosets of the kernel of the linear map ${\cal V}_z\mapsto{\cal Z}:~V_z\mapsto\Delta z=V_z^T\Delta\tilde{b}$ have the same number of elements. Let $d_z\leq n_z$ be the dimension of the range $\cal Z$ of this map. Then we have $2^{d_z}$ possible equiprobable $\Delta z$ for some fixed $\Delta\tilde{b}$. Only when $\Delta z=0$, which happens with probability $2^{-d_z}$, $\tilde{b}$ is not eliminated from $\T$ by the first measurement. The same reasoning can be done for a $\sigma_x$ measurement. Note that $d_z=d_x=0$ only holds for $\tilde{u}$ itself.

By performing the local Clifford operation and measurement on the $i$-th copy, a vector $\tilde{b}\in\Z_2^{nk}$ is transformed into $\bar{R}^T\tilde{b}\in\Z_2^{n(k-1)}$, where $\bar{R}$ is equal to $R$ without columns $(j-1)k+i$, for $j=1,\ldots, n$. For the second and each following measurement, the reasoning above can be repeated for the transformed $\bar{R}^T\tilde{b}$, except that we have $k-1,k-2,\ldots,k-m=\gamma k$ copies instead of $k$. A crucial observation is that for every next measurement, the probability that the state initially represented by $\tilde{b}$ is not eliminated, almost certainly remains the same during the entire process. Therefore, the probability that some $\tilde{b}$ for which $\cal Z$ has dimension $d_z$ and $\cal X$ has dimension $d_x$ is not eliminated after all measurements is equal to $2^{-k(d_z m_z+d_x m_x)}$. We postpone the proof to Appendix~\ref{appnext}.

\subsection{Minimal number of measurements}

So far we have given an information-theoretical interpretation of the protocol: we start with an unknown pure state (represented by $\tilde{u}$), which, with probability $\geq 1-\delta$, is contained in $\T$. Consecutive measurements rule out all inconsistent $\tilde{b}\in\T$. The probability that some $\tilde{b}\neq\tilde{u}$ survives this process is $2^{-k(d_z m_z+d_x m_x)}$. The total failure probability $p_F$ of the protocol is equal to $p_1+p_2$, where $p_1$ is the probability that $\tilde{u}\not\in\T$ in the first place and $p_2$ the probability that any $\tilde{b}\neq\tilde{u}$ survives the process. We already know that $p_1\leq\delta$. Now we calculate an upper bound for $p_2$ and the minimal fraction $m$ of all copies that has to be measured such that $p_F\rightarrow 0$ for $k\rightarrow\infty$.

To this end, we approximate the number of $\tilde{b}\in\T$ for which $\cal Z$ has dimension $\leq d_z$ and $\cal X$ has dimension $\leq d_x$. Call this number $N_{[d_z,d_x]}$. We will see that $N_{[d_z,d_x]}=2^{k[\alpha_{[d_z,d_x]}+O(k^{-1/4})]}$, where $\alpha_{[d_z,d_x]}>0$ is independent of $k$. Let $N_{[d_z,d_x]}^\ast=2^{k(\alpha_{[d_z,d_x]}^\ast+O(k^{-\eta}))}$ be the number of $\tilde{b}\in\T$ for which $\cal Z$ has dimension $=d_z$ and $\cal X$ has dimension $=d_x$, where $\eta>0$. Evidently,
\begin{equation}\label{sumN}
N_{[d_z,d_x]}=\sum_{d_z'\leq d_z,d_x'\leq d_x}N_{[d_z',d_x']}^\ast.
\end{equation}
The following inequality holds
\begin{eqnarray*}
p_2 &\leq& \sum_{[d_z,d_x]\neq [0,0]}^{[n_z,n_x]}N_{[d_z,d_x]}^\ast 2^{-k(d_z m_z+d_x m_x)} \\
 &=& \sum_{[d_z,d_x]\neq [0,0]}^{[n_z,n_x]}2^{-k[d_z m_z+d_x m_x-\alpha_{[d_z,d_x]}^\ast-O(k^{-\eta})]}.
\end{eqnarray*}
If we bound $m_z$ and $m_x$ by the following inequalities
\begin{equation}\label{ast}
d_z m_z+d_x m_x \geq \alpha_{[d_z,d_x]}^\ast+O(k^{-\zeta}),~\mbox{for all}~[d_z,d_x]\neq [0,0],
\end{equation}
where $0<\zeta<\eta$, it follows that $p_2\rightarrow 0$ for $k\rightarrow\infty$. Neglecting the vanishing terms, it can be verified that the inequalities
\begin{equation}\label{noast}
d_z m_z+d_x m_x \geq \alpha_{[d_z,d_x]}+O(k^{-1/2}),~\mbox{for all}~[d_z,d_x]\neq [0,0].
\end{equation}
are equivalent to (\ref{ast}). Indeed, it follows from (\ref{sumN}) that $\alpha_{[d_z,d_x]}=\alpha_{[d_z',d_x']}+O(k^{-1/4})=\alpha_{[d_z',d_x']}^\ast+O(k^{-1/4})$ for some $d_z'\leq d_z$ and $d_x'\leq d_x$. Since $d_z' m_z+d_x' m_x \geq \alpha_{[d_z',d_x']}^\ast=\alpha_{[d_z',d_x']}=\alpha_{[d_z,d_x]}$ (again neglecting vanishing terms) implies $d_z m_z+d_x m_x \geq \alpha_{[d_z,d_x]}$, a solution to (\ref{noast}) is also a solution to (\ref{ast}) and vice versa. From (\ref{noast}) and $N_{[d_z,d_x]}\geq N_{[d_z,d_x]}^\ast$, it follows that $p_2=O(2^{-\sqrt{k}})$.

This leaves us to calculate $N_{[d_z,d_x]}$. Let $G_z\in\Z_2^{n_z\times(n_z-d_z)}$ be a full rank matrix with column space ${\cal G}_z$. We define the space ${\cal W}_z({\cal G}_z)=\{V_zG_z~|~V_z\in{\cal V}_z\}$. Then all elements of ${\cal W}_z({\cal G}_z)^\perp=\{\Delta\tilde{b}\in\Z_2^{nk}~|~W_z^T\Delta\tilde{b}=0,~\forall W_z\in{\cal W}_z({\cal G}_z)\}$ correspond to a $\cal Z$ with dimension $\leq d_z$, as $G_z^T\Delta z=W_z^T\Delta\tilde{b}=0,~\forall \Delta z\in{\cal Z}$. We then have
\begin{equation*}
N_{[d_z,d_x]} = |\bigcup_{{\cal G}_z,{\cal G}_x}{\cal W}_z({\cal G}_z)^\perp \cap {\cal W}_x({\cal G}_x)^\perp \cap \T~|
\end{equation*}
where ${\cal G}_z$ and ${\cal G}_x$ run through all subspaces of $\Z_2^{n_z}$ and $\Z_2^{n_x}$ with dimension $n_z-d_z$ and $n_x-d_x$ respectively. It follows that
\begin{equation*}
N_{[d_z,d_x]} = r\max_{{\cal G}_z,{\cal G}_x}|{\cal W}_z({\cal G}_z)^\perp \cap {\cal W}_x({\cal G}_x)^\perp \cap \T|,
\end{equation*}
where $1\leq r\leq$ the total number of combinations $({\cal G}_z,{\cal G}_x)$, which is independent of $k$. Therefore, $r=O(1)$.

We now calculate $|{\cal W}_z({\cal G}_z)^\perp \cap {\cal W}_x({\cal G}_x)^\perp \cap \T|$. To this end, we first need to describe the spaces ${\cal W}_z({\cal G}_z)^\perp$, ${\cal W}_x({\cal G}_x)^\perp$ and their intersection in a simpler way. In the following, $e_t$ is a vector with a 1 on position $t$ and zeros elsewhere and $e$ is a vector with all ones. We investigate when $\Delta\tilde{b}\in{\cal W}_z(g)^\perp$, i.e. $(V_zg)^T\Delta\tilde{b}=0,~\forall V_z\in{\cal V}_z$, where $g\in\Z_2^{n_z}$. This can be written as
\begin{equation}\label{deltab}
\left[\begin{array}{c}g\otimes (A^{-1})_i\\\tilde{C}_x^T(\theta g\otimes e_i)\end{array}\right]^T\Delta\tilde{b}=0,
\end{equation}
for all possibilities of $(A^{-1})_i$ and $(C_l^T)_i$ ($l=n_z+1,\ldots, n$). It can be verified that
\[\tilde{C}_x^T(\theta g\otimes e_i)=(\theta g\otimes e)\odot c.\]
Therefore, (\ref{deltab}) is equivalent to
\[\left[\begin{array}{c}g\otimes a\\(\theta g\otimes e)\odot c\end{array}\right]^T\Delta\tilde{b}=0,\]
for all $a\in{\cal R}_a$ and $c\in{\cal R}_c$. Let $M_\theta$ be a matrix whose column space is the orthogonal complement of that of $L_\theta$. Then all possible $c$ are in the column space of $M_\theta\otimes I_k$. Since the distributions of $a$ and $c$ are independent, (\ref{deltab}) is equivalent to
\begin{equation}\label{Jz}
(\left[\begin{array}{cc}g & 0 \\ 0 & \theta ge^T\odot M_\theta\end{array}\right]^T\otimes I_k)~\Delta\tilde{b}=0.
\end{equation}
In an analogous way, we find that $\Delta\tilde{b}\in{\cal W}_x(g)^\perp$ iff
\begin{equation}\label{Jx}
(\left[\begin{array}{cc} 0 & \theta^T ge^T\odot M_{\theta^T} \\ g & 0\end{array}\right]^T\otimes I_k)~\Delta\tilde{b}=0.
\end{equation}
It is clear that $\Delta\tilde{b}\in{\cal W}_z({\cal G}_z)^\perp\cap{\cal W}_x({\cal G}_x)^\perp$ iff $\Delta\tilde{b}\in{\cal W}_z((G_z)_j)^\perp$, for $j=1\dots n_z-d_z$, and $\Delta\tilde{b}\in{\cal W}_x((G_x)_j)^\perp$, for $j=1\dots n_x-d_x$. We can write this as
\begin{equation*}
(J^T\otimes I_k)\Delta\tilde{b}=0,
\end{equation*}
where the column space $\cal J$ of $J$ is the sum of the column spaces of the matrices in (\ref{Jz}) over all $g=(G_z)_j$ and in (\ref{Jx}) over all $g=(G_x)_j$. This gives rise to the definition of $J$ given in section~\ref{sectionprotocol}.

We have found that $|{\cal W}_z({\cal G}_z)^\perp \cap {\cal W}_x({\cal G}_x)^\perp \cap \T|=$
\begin{equation*}
|\{\tilde{b}\in\T|(J^T\otimes I_k)\Delta\tilde{b}=0\}|.
\end{equation*}
Note that $(J^T\otimes I_k)\Delta\tilde{b}=0$ is equivalent to $(I_k\otimes J^T)\Delta\tilde{b}'=0$, or $J^T\Delta b_i=0$, for $i=1,\ldots, k$. The cosets $\Omega_j$ ($j=1,\ldots, q$) of the space $\Jt=\{w\in\Z_2^n|J^Tw=0\}$ constitute a partition of $\Z_2^n$. We want to know the number of $\tilde{b}\in\T$ for which $b_i$ is in the same coset as $u_i$, for all $i=1,\ldots, k$. In section~\ref{sectioninfo}, we derived that this number is equal to
\[2^{k[H-H_{\Jt}+O(\epsilon)]+O(\log_2k)}\]
\[\begin{array}[t]{cccl}
\mbox{where} & H &=& -\sum\limits_{b\in\Z_2^n}p(b)\log_2 p(b)\\
& H_{\Jt} &=& -\sum\limits_{j=1}^qp(\Omega_j)\log_2 p(\Omega_j).
\end{array}\]
Choose ${\cal G}_z$ (with dimension $n_z-d_z$) and ${\cal G}_x$ (with dimension $n_x-d_x$) such that $H_{\Jt}$ is minimal. We denote this minimum by $H_{[d_z,d_x]}$. Then it follows that
\begin{equation*}
N_{[d_z,d_x]}=2^{k[H-H_{[d_z,d_x]}+O(\epsilon)]+O(\log_2k)}.
\end{equation*}

Let $\epsilon=k^{-1/4}$. Then $p_1=\delta=O(k^{-1}\epsilon^{-2})=O(k^{-1/2})$. Recall that if (\ref{noast}) holds, $p_2=O(2^{-\sqrt{k}})$. Therefore, the probability $p_F$ that the protocol fails, is $O(k^{-1/2})$. Neglecting the vanishing terms, (\ref{noast}) can be formulated as the following linear programming problem:
\begin{equation*}\begin{array}{lcl}
\mbox{minimize} & & m=m_z+m_x \\
\mbox{subject to} & & d_z m_z+d_x m_x\geq H-H_{[d_z,d_x]}, \\
& & \mbox{for all}~[d_z,d_x]\neq[0,0],
\end{array}\end{equation*}
and we have $\gamma=(1-p_F)(1-m)\approx 1-m$. Note that, as $H\geq H_{[d_z,d_x]}$, the constraints where $d_x=0$ or $d_z=0$ of the LP problem imply that $m_z,m_x\geq 0$.

\section{An example}\label{sectionex}

In this section we illustrate the hashing protocol with an example. The 4-qubit cat state (also called GHZ state) is the state
\[\frac{1}{\sqrt{2}}(|0000\rangle+|1111\rangle)\]
which is stabilized by
\[\begin{array}{c}
\sigma_z\otimes I_2\otimes I_2\otimes\sigma_z\\
I_2\otimes\sigma_z\otimes I_2\otimes\sigma_z\\
I_2\otimes I_2\otimes\sigma_z\otimes\sigma_z\\
\sigma_x\otimes\sigma_x\otimes\sigma_x\otimes\sigma_x
\end{array}\]
and thus represented by
\[S_z=\left[\begin{array}{ccc}1&0&0\\0&1&0\\0&0&1\\1&1&1\end{array}\right],~S_x=\left[\begin{array}{c}1\\1\\1\\1\end{array}\right]~\mbox{and}~b=\left[\begin{array}{c}0\\0\\0\\0\end{array}\right].\]
It is straightforward that nothing is gained by measuring according to a partition other than exclusively $\sigma_z$ measurements or $\sigma_x$ measurements. With (\ref{theta}), we have $\theta=[1~1~1]$. Note that $\theta$ is not orthogonal. We find $L_\theta=1$ and $L_{\theta^T}=[0~0~0]^T$. The linear constraints (\ref{B'})-(\ref{C'}) become
\[\begin{array}{c}
B_1+B_2+B_3+B_4=0\\
C_1=C_2=C_3=C_4=0
\end{array}\]
so a local Clifford operation that results in a permutation of all possible $\tilde{b}$ is of the form
\[\left[\begin{array}{cccc|cccc}
A &  & & & B_1 & &  &  \\
 & A &  & & & B_2 & &  \\
 &  & A & & &  & B_3 & \\
 &  & & A &  &  & & B_1+B_2+B_3 \\ \hline
 &  &  & & A^{-T} & & &  \\
 &  &  & &  & A^{-T} & &  \\
 &  &  & & &  & A^{-T} & \\
 &  &  & & &  & & A^{-T} \\
\end{array}\right]\]
and $R$ is of the form
\[\left[\begin{array}{cccc}
A^{-1} & & & B_1^T \\
 & A^{-1} & & B_2^T \\
 & & A^{-1}  & B_3^T \\
 & & & A^T
\end{array}\right].\]

We formulate the linear programming problem to calculate the yield of the protocol. At the start, the 4 parties share $k$ copies of a state
\[\rho=\sum\limits_{b\in\Z_2^4}p_b|\psi_b\rangle\langle\psi_b|,\quad\mbox{where}\]
$|\psi_b\rangle=\frac{1}{\sqrt{2}}(|b_1,b_2,b_3,0\rangle+(-1)^{b_4}|b_1+1,b_2+1,b_3+1,1\rangle)$.
From $L_\theta, L_{\theta^T}$ we find $M_\theta=0$ and $M_{\theta^T}=I_3$. We now calculate $H_{[d_z,d_x]}$ for different values of $d_z,d_x$. When $d_x=0$, we have $G_x=1$ and $V=I_3$. It follows that $\Jt=\{0\}$ and therefore $H_{[d_z,0]}=H$, for all $d_z>0$.
When $d_x=1$, we have $G_x=0$ and $V=0$. From $M_\theta=0$, it follows that $U=0$. We now have
\[J=\left[\begin{array}{c}G_z\\0\end{array}\right].\]
Evidently, $H_{[3,1]}=H_{[n_z,n_x]}=0$. When $d_z=0$, we have $G_z=I_3$. It follows that
\[H_{[0,1]}=-\sum_{b_{123}\in\Z_2^3}(\sum_{b_4\in\Z_2}p_b)\log_2(\sum_{b_4\in\Z_2}p_b).\]
In both cases $d_z=1$ and $d_z=2$, we have to calculate $H_{\Jt}$ for seven different subspaces $\Jt$. The minimum is $H_{[1,1]}$ or $H_{[2,1]}$ respectively. As an example, let $d_z=1$ and
\[G_z=\left[\begin{array}{cc}1&0\\1&1\\0&1\end{array}\right].\]
The four cosets of $\Jt$ are then (the first column is $\Jt$):
\[\begin{array}{c|c|c|c}
0000 & 0010 & 0100 & 1000 \\
0001 & 0011 & 0101 & 1001 \\
1110 & 1100 & 1010 & 0110 \\
1111 & 1101 & 1011 & 0111
\end{array}\]
The LP problem is now
\begin{equation*}\begin{array}{lcl}
\mbox{minimize} & & m=m_z+m_x \\
\mbox{subject to} & & m_z\geq 0 \\
 & & m_x\geq H-H_{[0,1]} \\
 & & m_z+m_x\geq H-H_{[1,1]} \\
 & & 2m_z+m_x\geq H-H_{[2,1]} \\
 & & 3m_z+m_x\geq H. \\
\end{array}\end{equation*}

\begin{figure}
\includegraphics[width=0.45\textwidth]{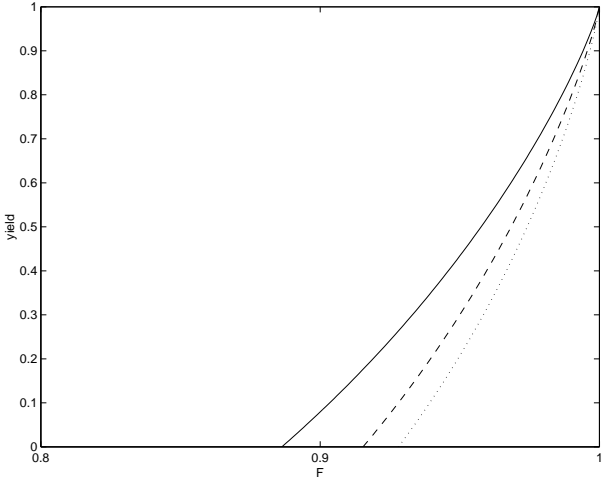}
\caption{\label{yield} comparison of different protocols for the given cat state example. The dotted line gives the yield of the protocol of Ref.~\cite{man}, the dashed line of that of Ref.~\cite{lo} and the solid line of our protocol, as a function of the fidelity $F$ of the depolarizing channels.}
\end{figure}

For this example, we have compared our protocol to those of Refs.~\cite{man,lo}. We start with copies of the 4-qubit cat state, prepared by the first party. The second, third and fourth qubit of each copy is sent through identical depolarizing channels to the corresponding parties. The action of each channel is
\[\rho\mapsto F\rho+\frac{1-F}{3}(\sigma_x\rho\sigma_x^{\dag}+\sigma_y\rho\sigma_y^{\dag}+\sigma_z\rho\sigma_z^{\dag}).\]
and we call $F$ the fidelity of the channels. It can be verified that this yields a mixture with probabilities:
\[\left[\begin{array}{c}
p_{0000} \\ p_{0001} \\ p_{0010} \\ p_{0011} \\ p_{0100} \\ p_{0101} \\ p_{0110} \\ p_{0111} \\ p_{1000} \\ p_{1001} \\ p_{1010} \\ p_{1011} \\ p_{1100} \\ p_{1101} \\ p_{1110} \\ p_{1111}\end{array}\right] =
\left[\begin{array}{cccc}
     1   &  0  &   3  &   0 \\
     0   &  3  &   0  &   1 \\
     0   &  1  &   2  &   1 \\
     0   &  1  &   2  &   1 \\
     0   &  1  &   2  &   1 \\
     0   &  1  &   2  &   1 \\
     0   &  0  &   2  &   2 \\
     0   &  0  &   2  &   2 \\
     0   &  0  &   0  &   4 \\
     0   &  0  &   0  &   4 \\
     0   &  0  &   2  &   2 \\
     0   &  0  &   2  &   2 \\
     0   &  0  &   2  &   2 \\
     0   &  0  &   2  &   2 \\
     0   &  1  &   2  &   1 \\
     0   &  1  &   2  &   1
\end{array}\right]
\left[\begin{array}{c} F^3 \\ F^2\frac{1-F}{3} \\ F\left(\frac{1-F}{3}\right)^2 \\ \left(\frac{1-F}{3}\right)^3
\end{array}\right].\]
The yield of our protocol for this example is plotted as a function of the fidelity of the channels in figure~\ref{yield}. So is the yield of the protocol of Ref.~\cite{man}:
\[1-\max\limits_{j=1,2,3}[H(b_j)]-H(b_4)\]
and the yield of the improved protocol of Ref.~\cite{lo}:
\[\begin{array}{rl}
\max & \left(1-\max\limits_{j=1,2,3}[H(b_j)]-H(b_4|b_1,b_2,b_3), \right. \\ & \left. 1-\max\limits_{j=1,2,3}[H(b_j|b_4)]-H(b_4)\right).
\end{array}\]

Finally, we mention that for every cat state, it can be verified that there is no benefit in using more general local Clifford operations than CNOTs. 
We give another example where not only applying CNOTs pays off. Suppose we want to distill copies of the 8-qubit CSS state represented by
\[\theta=\left[\begin{array}{cccc}
0&1&1&1\\
1&0&1&1\\
1&1&0&1\\
1&1&1&0\\
\end{array}\right].\]
Note that, as $\theta$ is orthogonal, $R$ is given by (\ref{RO}). The intial mixed states are diagonal in the $S$-basis, with probabilities $p_0=3/4$, $p_{1 b_{2\ldots 8}}=0$, for all $b_{2\ldots 8}\in\Z_2^7$, and $p_{0 b_{2\ldots 8}}=1/[4(2^7-1)]$, for all $b_{2\ldots 8}\neq 0\in\Z_2^7$. It can now be verified that the yield of our hashing protocol is equal to 
\[\gamma=1-\frac{H}{4}\approx 0.36.\]
Applying only CNOTs, the yield is equal to
\begin{eqnarray*}
\lefteqn{1-\frac{H(b_{5\ldots 8})}{4}-\frac{H-H(b_{5\ldots 8})}{3} \approx 0.29} & & \\
&=& 1-\frac{H}{4}-\frac{H-H(b_{5\ldots 8})}{12} \\
&<& 1-\frac{H}{4}=\gamma.
\end{eqnarray*}

\section{Conclusion}

We have presented a hashing protocol to distill multipartite CSS states, an important class of stabilizer states. Starting with $k$ copies of a mixed state that is diagonal in the $S$-basis, the protocol consists of local Clifford operations that result in a permutation of all $2^{nk}$ tensor products of CSS states, followed by Pauli measurements that extract information on the global state. To find these local Clifford operations, we used the efficient binary matrix description of stabilizer states and Clifford operations. With the aid of the information-theoretical notion of a strongly typical set, it is possible to calculate the minimal number of copies that have to be measured in order to end up with copies of a pure CSS state, for $k$ approaching infinity. As a result, the yield of the protocol is formulated as the solution of a linear programming problem.

\appendix

\section{Solving Eqs.~(\ref{A})-(\ref{C})}\label{appclif}

First, we show that (\ref{IA}) follows from (\ref{A})-(\ref{D}). Comparing each corresponding block on both sides of (\ref{A}) yields:
\[A_v=A_{n_z+u}\quad\mbox{if}~\theta_{uv}=1,~\mbox{for}~u=1,\ldots, n_x~\mbox{and}~v=1,\ldots, n_z.\]
From this, it is clear that all $A_i$ ($i=1,\ldots, n$) must be equal. If not, it is possible to divide $\{1,\ldots,n\}$ into two disjunct nonempty subsets $\omega_1$ and $\omega_2$ for which $\theta_{uv}=0$ if $n_z+u\in\omega_1$ and $v\in\omega_2$ or vice versa. We could permute rows and columns of $\theta$ such that the resulting $\theta'=T_r\theta T_c$ has all rows $u_1$ for which $n_z+u_1\in\omega_1$ above rows $u_2$ for which $n_z+u_2\in\omega_2$, and all columns $v_1$ for which $v_1\in\omega_1$ on the left of columns $v_2$ for which $v_2\in\omega_2$. We then have
\[\left[\begin{array}{cc}T_c^T & 0\\0 & T_r\end{array}\right]\left[\begin{array}{c}I\\\theta\end{array}\right]T_c=\left[\begin{array}{c}I\\\theta'\end{array}\right]=\left[\begin{array}{cc}I & 0\\0 & I\\\ast & 0\\0 & \ast\end{array}\right].\]
It is clear that this represents a separable CSS state, which we excluded from the beginning. An analogous proof holds for the $D_i$.

Second, we show that if $\theta$ is not orthogonal, with (\ref{B})-(\ref{C}) we can find subsets $Z_B$ and $Z_C$ of $\{1,\ldots,n\}$ for which all $B_i$ and $C_i$ are zero if $i\in Z_B$ or $Z_C$ respectively. Note that (\ref{B}) is equivalent to
\begin{equation*}
(S_x^T\otimes I_k)\tilde{B}(S_x\otimes I_k)=0.
\end{equation*}
We can rewrite this as linear constraints on the $B_i$ as follows
\begin{equation}\label{Lx}
(L_x^T\otimes I_k)\left[\begin{array}{c}B_1\\\vdots\\B_n\end{array}\right]=0.
\end{equation}
The $n$-bit columns of $L_x$ are $(S_x)_j\odot(S_x)_l,~\forall j,l: 1\leq j\leq l\leq n_x$. Note that (\ref{Lx}) is the same as (\ref{B'}). We can do the same for (\ref{C}). We denote the column spaces of $L_x$ and $L_z$ by ${\cal L}_x$ and ${\cal L}_z$ respectively. As the constraints (\ref{B})-(\ref{C}) are independent, all solutions $\tilde{B}$ must be consistent with all solutions $\tilde{C}$. From (\ref{B})-(\ref{C}), it follows that
\[\begin{array}{rcl}
(\theta\otimes I_k)\tilde{B}_z\tilde{C}_z &=& (\theta\otimes I_k)\tilde{B}_z(\theta^T\otimes I_k)\tilde{C}_x(\theta\otimes I_k) \\
 &=& \tilde{B}_x\tilde{C}_x(\theta\otimes I_k).
\end{array}\]
In the same way as for (\ref{IA}), we can prove then that $B_1C_1=\ldots=B_nC_n$. If $B_iC_i=0$, then either $B_i=0$ or $C_i=0$. Indeed, suppose $B_i\neq 0$. Then $e_i\not\in{\cal L}_x$. Consequently, there exist some solution $p$ to $L_x^Tp=0$ with $(p)_i=1$. Note that $p\otimes I_k$ is a solution to (\ref{Lx}). It follows that $B_iC_i=I_kC_i=0$.

This leaves us to prove that $B_iC_i\neq 0$ only if $\theta$ is orthogonal. Suppose $B_iC_i\neq 0$, for all $i=1,\ldots, n$, then, for every $i$, there exists a solution $p$ to $L_x^Tp=0$ with $(p)_i=1$. It is clear that, for every $i$ and $j$, there also exists a solution $p$ with $(p)_i=(p)_j=1$. So, for every $i$ and $j$, we have a solution $\tilde{B}$ to (\ref{Lx}) with $B_i=B_j=I_k$ that must be consistent with all solutions $\tilde{C}$. It follows that $C_1=\ldots=C_n$. The same holds for the $B_i$. This implies that the spaces ${\cal L}_x$ and ${\cal L}_z$ are equal and consist of all vectors of even weight. No vector of odd weight is in ${\cal L}_z$, otherwise ${\cal L}_z$ would be the entire space $\Z_2^n$ and consequently $C_i=0$. So all $(S_x)_j\odot(S_x)_l$ and $(S_z)_j\odot(S_z)_l$ must have even weight. With (\ref{theta}), it can be verified that this only holds if $(\theta)_u^T(\theta)_v=(\theta^T)_u^T(\theta^T)_v=\delta_{uv}$, where $\delta_{uv}$ is the Kronecker delta. This is equivalent with $\theta^T\theta=\theta\theta^T=I$.

\section{Proof of constant elimination probability}\label{appnext}

We show that the probability that a state, initially represented by $\tilde{b}$ for which $\cal Z$ has dimension $d_z$ and $\cal X$ has dimension $d_x$, is not eliminated after the protocol has ended, is equal to $2^{-k(d_zm_z+d_xm_x)+O(2^{-\gamma k})}$. First, we show that this probability $\geq 2^{-k(d_zm_z+d_xm_x)}$. Without loss of generality, we assume that the $i$-th copy is measured in the $i$-th step. We consider all measurements performed at the end (cfr. the two equivalent views of the protocol depicted in figure~\ref{eqviews}) and we call the overall transformation matrix $R$. Then the $i$-th measurement in fact reveals $(R^T\tilde{u})_{(j-1)k+i}$, for $j=1,\ldots, n_z$, if it is a $\sigma_z$ measurement or for $j=n_z+1,\ldots, n$ if it is a $\sigma_x$ measurement. Following the reasoning of section~\ref{sectionyield}, it is clear that for each measurement, no other outcome $\Delta z$ or $\Delta x$ than those in ${\cal Z}$ or in ${\cal X}$ can occur.

However, it is possible that during the process (after some measurements), one or both of the sets of outcomes $\bar{\cal Z}$ and $\bar{\cal X}$ (corresponding to the transformed $\bar{R}^T\tilde{b}$) are strictly smaller than $\cal Z$ and $\cal X$, which means that the probability of not being eliminated by a measurement is larger than at the start. Suppose the first measurement is a $\sigma_z$ measurement on the $k$-th copy. Recall that a measurement inevitably involves the loss of the phase factors of observables noncommuting with the measurement. This loss of information causes initially different $\tilde{b}\in\Z_2^{nk}$ to be mapped to the same vector in $\Z_2^{n(k-1)}$. Indeed, $\tilde{b}$ is mapped to $\bar{R}^T\tilde{b}$, where $\bar{R}$ is equal to $R$ without columns $jk$, for $j=1,\ldots, n$. We investigate when $\bar{R}^T\tilde{v}=\bar{R}^T\tilde{w}$ and $\tilde{v}, \tilde{w}$ correspond concerning the measurement outcome (otherwise at most one is not eliminated). This is the case iff $(R^T)_l(\tilde{v}+\tilde{w})=0$, for all $l$ except $(n_z+j)k$, for $j=1,\ldots, n_x$. Equivalently, $\tilde{v}+\tilde{w}\in{\cal Q}$, where ${\cal Q}$ is the $n_x$-dimensional space generated by columns $(n_z+j)k$, for $j=1,\ldots, n_x$, of $R^{-T}$. If we assume that $\theta$ is not orthogonal (the orthogonal case is analogous), then from (\ref{Req}) and (\ref{RN}), we have
\begin{equation*}
R^{-T}=\left[\begin{array}{cc}
I_{n_z}\otimes A^T & (\theta^T\otimes I_k)\tilde{C}_x^T \\
(\theta\otimes I_k)\tilde{B}_z^T & I_{n_x}\otimes A^{-1}
\end{array}\right].
\end{equation*}

Let $\Jt$ be defined as in section~\ref{sectionprotocol}, where ${\cal G}_z$ and ${\cal G}_x$ have dimensions $n_z-d_z'$ and $n_x-d_x'$ respectively and $d_z'<d_z$ or $d_x'<d_x$. Consequently, $\Delta\tilde{b}\not\in\Jt\otimes\Z_2^k$. We investigate when $\bar{R}^T\tilde{b}\in\Jt\otimes\Z_2^{k-1}$. For every $\Delta\tilde{v}\in\Z_2^{n(k-1)}$ that satisfies $\Delta v_i\in\Jt$, for $i=1,\ldots, k-1$, there is a $\Delta\tilde{w}\in\Z_2^{nk}$ that satisfies $\Delta w_i\in\Jt$, for $i=1,\ldots, k$, and $\bar{R}^T\Delta\tilde{w}=\Delta\tilde{v}$. Indeed, define some $\Delta\tilde{t}\in\Z_2^{nk}$ such that $\Delta t_i=\Delta v_i$, for $i=1,\ldots, k-1$, and $\Delta t_k\in\Jt$. Let $\Delta\tilde{w}=R^{-T}\Delta\tilde{t}$. From the definition of $\bar{R}$, it follows that $\bar{R}^T\Delta\tilde{w}=\Delta\tilde{v}$. In the previous paragraph, we have shown that the set of all $\Delta\tilde{t}$ that satisfy $\Delta t_i\in\Jt$ is invariant under left multiplication by some $R^T$, where $R$ is given by (\ref{RN}). As $R$ is invertible, the same holds for $R^{-T}$. Therefore, $\Delta w_i\in\Jt$, for $i=1,\ldots, k$. It follows that $\bar{R}^T\tilde{b}\in\Jt\otimes\Z_2^{k-1}$ iff there is some $\tilde{q}\in{\cal Q}$ and some $\Delta\tilde{w}\in\Jt\otimes\Z_2^k$ such that $\Delta\tilde{b}+\tilde{q}=\Delta\tilde{w}$.

Let $\tilde{q}(v)=\sum_j (v)_j(R^{-T})_{(n_z+j)k}\in{\cal Q}$, where $v\in\Z_2^{n_x}$. In the same way as in section~\ref{sectionprotocol}, it can be verified that $q_i(v)$, for $i=1,\ldots, k$, all satisfy the same linear constraints. Let ${\cal L}_v$ be the space of vectors that satisfy these constraints. All $q_i(v)$, for $i=1,\ldots, k$, are uniformly and independently distributed over ${\cal L}_v$. If ${\cal L}_v\subset\Jt$, then there is no $\tilde{q}(v)$ such that $\Delta b_i+q_i(v)\in\Jt$, as $\Delta b_i\not\in\Jt$ for some $i$. Therefore, ${\cal L}_v$ must $\not\subset\Jt$. Let $l\geq 2$ be the number of cosets ${\cal L}_v\cap\Jt$ within ${\cal L}_v$. All cosets have the same number of elements. Therefore, the probability that $\Delta b_i+q_i(v)\in\Jt$ is at most $l^{-1}\leq 2^{-1}$. Note that if $(\Delta b_i+\Jt)\cap{\cal L}_v=\emptyset$, this probability is zero. Because $q_i(v)$, for $i=1,\ldots, k$, are independent, the probability that $\Delta b_i+q_i(v)\in\Jt$, for all $i=1,\ldots, k$, is at most $2^{-k}$. The probability that there is some $\tilde{q}\in{\cal Q}$ such that $\Delta b_i+q_i\in\Jt$, for all $i=1,\ldots, k$, is then at most $2^{-k+n_x}$. The probability that $|\bar{\cal Z}|<|{\cal Z}|$ or $|\bar{\cal X}|<|{\cal X}|$ after the last measurement of the protocol, is therefore at most
\[r\sum_{t=1}^{mk}2^{-(k-t)+n}<rmk2^{-\gamma k+n}=\xi,\]
where $r$, independent of $k$, is the total number of combinations $({\cal G}_z,{\cal G}_x)$ with proper dimensions. Note that $\xi=O(2^{-\gamma k})$. The probability that $\tilde{b}$ is not eliminated by a $\sigma_z$ (or $\sigma_x$) measurement is at most $2^{-d_z}+\xi$ (or $2^{-d_x}+\xi$). Consequently, the probability that $\tilde{b}$ survives the entire process is at most
\[(2^{-d_z}+\xi)^{m_zk}(2^{-d_x}+\xi)^{m_xk}=2^{-k(d_zm_z+d_xm_x)+O(2^{-\gamma k})}.\]

\begin{acknowledgments}
We thank Maarten Van den Nest for interesting discussions.
Research funded by a Ph.D. grant of the Institute for the Promotion of Innovation through Science and Technology in Flanders (IWT-Vlaanderen).
Dr. Bart De Moor is a full professor at the Katholieke Universiteit Leuven, Belgium. Research supported by Research Council KUL: GOA AMBioRICS, CoE EF/05/006 Optimization in Engineering, several PhD/postdoc \& fellow grants; Flemish Government: FWO: PhD/postdoc grants, projects, G.0407.02 (support vector machines), G.0197.02 (power islands), G.0141.03 (Identification and cryptography), G.0491.03 (control for intensive care glycemia), G.0120.03 (QIT), G.0452.04 (new quantum algorithms), G.0499.04 (Statistics), G.0211.05 (Nonlinear), G.0226.06 (cooperative systems and optimization), G.0321.06 (Tensors), G.0553.06 (VitamineD), research communities (ICCoS, ANMMM, MLDM); IWT: PhD Grants,GBOU (McKnow), Eureka-Flite2; Belgian Federal Science Policy Office: IUAP P5/22 ('Dynamical Systems and Control: Computation, Identification and Modelling', 2002-2006) ; PODO-II (CP/40: TMS and Sustainability); EU: FP5-Quprodis; ERNSI; Contract Research/agreements: ISMC/IPCOS, Data4s, TML, Elia, LMS, Mastercard.
\end{acknowledgments}


\begin{thebibliography}{4}
\expandafter\ifx\csname natexlab\endcsname\relax\def\natexlab#1{#1}\fi
\expandafter\ifx\csname bibnamefont\endcsname\relax
  \def\bibnamefont#1{#1}\fi
\expandafter\ifx\csname bibfnamefont\endcsname\relax
  \def\bibfnamefont#1{#1}\fi
\expandafter\ifx\csname citenamefont\endcsname\relax
  \def\citenamefont#1{#1}\fi
\expandafter\ifx\csname url\endcsname\relax
  \def\url#1{\texttt{#1}}\fi
\expandafter\ifx\csname urlprefix\endcsname\relax\def\urlprefix{URL }\fi
\providecommand{\bibinfo}[2]{#2}
\providecommand{\eprint}[2][]{\url{#2}}

\bibitem{D:03}
\bibinfo{author}{\bibfnamefont{J.}~\bibnamefont{Dehaene}}
  \bibnamefont{and}
  \bibinfo{author}{\bibfnamefont{B.}~\bibnamefont{{De Moor}}},
  \emph{\bibinfo{title}{The Clifford group, stabilizer states, and linear and quadratic operations over GF(2)}},
  \bibinfo{journal}{Phys. Rev. A} \textbf{\bibinfo{volume}{68}},
  \bibinfo{pages}{042318} (\bibinfo{year}{2003}).

\bibitem{asch}
\bibinfo{author}{\bibfnamefont{H.}~\bibnamefont{Aschauer}},
\bibinfo{author}{\bibfnamefont{W.}~\bibnamefont{D\"ur}}
  \bibnamefont{and}
  \bibinfo{author}{\bibfnamefont{H.-J.}~\bibnamefont{Briegel}},
  \emph{\bibinfo{title}{Multiparticle entanglement purification for two-colorable graph states}},
  \bibinfo{journal}{Phys. Rev. A} \textbf{\bibinfo{volume}{71}},
  \bibinfo{pages}{012319} (\bibinfo{year}{2005}).

\bibitem{dur1}
\bibinfo{author}{\bibfnamefont{W.}~\bibnamefont{D\"ur}},
\bibinfo{author}{\bibfnamefont{H.}~\bibnamefont{Aschauer}} 
\bibnamefont{and}
  \bibinfo{author}{\bibfnamefont{H.-J.}~\bibnamefont{Briegel}},
  \emph{\bibinfo{title}{Multiparticle entanglement purification for graph states}},
  \bibinfo{journal}{Phys. Rev. Lett.} \textbf{\bibinfo{volume}{91}},
  \bibinfo{pages}{107903} (\bibinfo{year}{2003}).

\bibitem{man}
\bibinfo{author}{\bibfnamefont{E.N.}~\bibnamefont{Maneva}}
  \bibnamefont{and}
  \bibinfo{author}{\bibfnamefont{ J.A.}~\bibnamefont{Smolin}},
  \emph{\bibinfo{title}{Improved two-party and multi-party purification protocols}},
  \eprint{quant-ph/0003099}.

\bibitem{lo}
\bibinfo{author}{\bibnamefont{{Kai Chen}}}
  \bibnamefont{and}
  \bibinfo{author}{\bibnamefont{{Hoi-Kwong Lo}}},
  \emph{\bibinfo{title}{Multi-partite quantum cryptographic protocols with noisy {GHZ} states}},
  \eprint{quant-ph/0404133}.

\bibitem{Bennett}
\bibinfo{author}{\bibfnamefont{C.H.}~\bibnamefont{Bennett}},
\bibinfo{author}{\bibfnamefont{D.P.}~\bibnamefont{DiVincenzo}},
\bibinfo{author}{\bibfnamefont{J.A.}~\bibnamefont{Smolin}}
  \bibnamefont{and}
  \bibinfo{author}{\bibfnamefont{W.K.}~\bibnamefont{Wootters}},
  \emph{\bibinfo{title}{Mixed-state entanglement and quantum error correction}},
  \bibinfo{journal}{Phys. Rev. A} \textbf{\bibinfo{volume}{54}},
  \bibinfo{pages}{3824} (\bibinfo{year}{1996}).

\bibitem{DVD:03}
\bibinfo{author}{\bibfnamefont{J.}~\bibnamefont{Dehaene}},
  \bibinfo{author}{\bibfnamefont{M.}~\bibnamefont{{Van den Nest}}},
  \bibinfo{author}{\bibfnamefont{B.}~\bibnamefont{{De Moor}}},
  \bibnamefont{and}
  \bibinfo{author}{\bibfnamefont{F.}~\bibnamefont{Verstraete}},
  \emph{\bibinfo{title}{Local permutations of products of {Bell} states and entanglement distillation}},
  \bibinfo{journal}{Phys. Rev. A} \textbf{\bibinfo{volume}{67}},
  \bibinfo{pages}{022310} (\bibinfo{year}{2003}).

\bibitem{CT}
\bibinfo{author}{\bibfnamefont{T.M.}~\bibnamefont{Cover}}
  \bibnamefont{and}
  \bibinfo{author}{\bibfnamefont{J.A.}~\bibnamefont{Thomas}},
  \emph{\bibinfo{title}{Elements of Information Theory}},
  \bibinfo{publisher}{John Wiley \& Sons, Inc.} (\bibinfo{year}{1991}).

\bibitem{cheb}
\bibinfo{author}{\bibfnamefont{E.W.}~\bibnamefont{Weisstein}},
  \emph{\bibinfo{title}{Chebyshev inequality}},
  \bibinfo{publisher}{From MathWorld - A Wolfram Web Resource},
  \eprint{mathworld.wolfram.com/ChebyshevInequality.html}.

\bibitem{stir}
\bibinfo{author}{\bibfnamefont{E.W.}~\bibnamefont{Weisstein}},
  \emph{\bibinfo{title}{Stirling's approximation}},
  \bibinfo{publisher}{From MathWorld - A Wolfram Web Resource},
  \eprint{mathworld.wolfram.com/StirlingsApproximation.html}.

\bibitem{R:03}
\bibinfo{author}{\bibfnamefont{R.}~\bibnamefont{Raussendorf}},
\bibinfo{author}{\bibfnamefont{D.E.}~\bibnamefont{Browne}}
  \bibnamefont{and}
  \bibinfo{author}{\bibfnamefont{H.J.}~\bibnamefont{Briegel}},
  \emph{\bibinfo{title}{Measurement-based quantum computation with cluster states}},
  \bibinfo{journal}{Phys. Rev. A} \textbf{\bibinfo{volume}{68}},
  \bibinfo{pages}{022312} (\bibinfo{year}{2003}).

\bibitem{GPhD}
\bibinfo{author}{\bibfnamefont{D.}~\bibnamefont{Gottesman}},
  \emph{\bibinfo{title}{Stabilizer codes and quantum error correction}},
  \bibinfo{journal}{Caltech Ph.D. thesis},
  \eprint{quant-ph/9705052}.

\bibitem{G:98}
\bibinfo{author}{\bibfnamefont{D.}~\bibnamefont{Gottesman}},
  \emph{\bibinfo{title}{A Theory of Fault-Tolerant Quantum Computation}},
  \bibinfo{journal}{Phys. Rev. A} \textbf{\bibinfo{volume}{57}},
  \bibinfo{pages}{127} (\bibinfo{year}{1998}).

\bibitem{B2}
\bibinfo{author}{\bibfnamefont{C.H.}~\bibnamefont{Bennett}},
\bibinfo{author}{\bibfnamefont{G.}~\bibnamefont{Brassard}},
\bibinfo{author}{\bibfnamefont{C.}~\bibnamefont{Cr\'epeau}},
\bibinfo{author}{\bibfnamefont{R.}~\bibnamefont{Josza}},
\bibinfo{author}{\bibfnamefont{A.}~\bibnamefont{Peres}}
\bibnamefont{and}
\bibinfo{author}{\bibfnamefont{W.K.}~\bibnamefont{Wootters}},
  \emph{\bibinfo{title}{Teleporting an unknown quantum state via dual classical and Einstein-Podolsky-Rosen channels}},
  \bibinfo{journal}{Phys. Rev. Lett.} \textbf{\bibinfo{volume}{70}},
  \bibinfo{pages}{1895} (\bibinfo{year}{1993}).

\bibitem{B3}
\bibinfo{author}{\bibfnamefont{C.H.}~\bibnamefont{Bennett}}
\bibnamefont{and}
\bibinfo{author}{\bibfnamefont{S.J.}~\bibnamefont{Wiesner}},
  \emph{\bibinfo{title}{Communication via one- and two-particle operators on Einstein-Podolsky-Rosen states}},
  \bibinfo{journal}{Phys. Rev. Lett.} \textbf{\bibinfo{volume}{69}},
  \bibinfo{pages}{2881} (\bibinfo{year}{1992}).

\bibitem{ekert}
\bibinfo{author}{\bibfnamefont{A.}~\bibnamefont{Ekert}},
  \emph{\bibinfo{title}{Quantum cryptography based on Bell's theorem}},
  \bibinfo{journal}{Phys. Rev. Lett.} \textbf{\bibinfo{volume}{67}},
  \bibinfo{pages}{661} (\bibinfo{year}{1991}).

\bibitem{dur2}
\bibinfo{author}{\bibfnamefont{W.}~\bibnamefont{D\"ur}},
\bibinfo{author}{\bibfnamefont{J.}~\bibnamefont{Calsamiglia}}
  \bibnamefont{and}
  \bibinfo{author}{\bibfnamefont{H.-J.}~\bibnamefont{Briegel}},
  \emph{\bibinfo{title}{Multipartite secure state distribution}},
  \bibinfo{journal}{Phys. Rev. A} \textbf{\bibinfo{volume}{71}},
  \bibinfo{pages}{042336} (\bibinfo{year}{2005}).

\bibitem{karlsson}
\bibinfo{author}{\bibfnamefont{A.}~\bibnamefont{Karlsson}},
\bibinfo{author}{\bibfnamefont{M.}~\bibnamefont{Koashi}}
\bibnamefont{and}
\bibinfo{author}{\bibfnamefont{N.}~\bibnamefont{Imoto}},
  \emph{\bibinfo{title}{Quantum entanglement for secret sharing and secret splitting}},
  \bibinfo{journal}{Phys. Rev. A} \textbf{\bibinfo{volume}{59}},
  \bibinfo{pages}{162} (\bibinfo{year}{1999}).

\bibitem{hillery}
\bibinfo{author}{\bibfnamefont{M.}~\bibnamefont{Hillery}},
\bibinfo{author}{\bibfnamefont{V.}~\bibnamefont{Buzek}}
\bibnamefont{and}
\bibinfo{author}{\bibfnamefont{A.}~\bibnamefont{Berthiaume}},
  \emph{\bibinfo{title}{Quantum secret sharing}},
  \bibinfo{journal}{Phys. Rev. A} \textbf{\bibinfo{volume}{59}},
  \bibinfo{pages}{1829} (\bibinfo{year}{1999}).

\bibitem{cleve}
\bibinfo{author}{\bibfnamefont{R.}~\bibnamefont{Cleve}},
\bibinfo{author}{\bibfnamefont{D.}~\bibnamefont{Gottesman}}
\bibnamefont{and}
\bibinfo{author}{\bibnamefont{Hoi-Kwong Lo}},
  \emph{\bibinfo{title}{How to share a quantum secret}},
  \bibinfo{journal}{Phys. Rev. Lett.} \textbf{\bibinfo{volume}{83}},
  \bibinfo{pages}{648} (\bibinfo{year}{1999}).

\bibitem{crep}
\bibinfo{author}{\bibfnamefont{C.}~\bibnamefont{Cr\'epeau}},
\bibinfo{author}{\bibfnamefont{D.}~\bibnamefont{Gottesman}}
\bibnamefont{and}
\bibinfo{author}{\bibfnamefont{A.}~\bibnamefont{Smith}},
  \emph{\bibinfo{title}{Secure Multi-party Quantum Computing}},
  \bibinfo{journal}{Proc. STOC 2002},
  \eprint{quant-ph/0206138}.

\end{thebibliography}
\end{document}